\pdfoutput=1
\newcommand*{\ATLASLATEXPATH}{}
\documentclass[PAPER, atlasdraft=false, texlive=2016, UKenglish, cernpreprint]{\ATLASLATEXPATH atlasdoc}
 
\usepackage{\ATLASLATEXPATH atlaspackage}
\usepackage[utf8]{inputenc}
\usepackage[backref=false]{\ATLASLATEXPATH atlasbiblatex}
 
\usepackage{\ATLASLATEXPATH atlasphysics}
 
\graphicspath{{logos/}{figures/}}

\AtlasTitle{Search for high-mass dilepton resonances using 139~\ifb\ of $pp$ collision data collected at $\sqrt{s}=13$~\TeV\ with the ATLAS detector }
 
\AtlasVersion{3.0}
\AtlasJournalRef{Phys. Lett. B 796 (2019) 68}
\AtlasDOI{10.1016/j.physletb.2019.07.016}
 
\AtlasAbstract{
A search for high-mass dielectron and dimuon resonances in the mass range of 250$\,$\GeV\ to 6$\,$\TeV\ is presented.
The data were recorded by the ATLAS experiment in proton--proton collisions at a centre-of-mass energy of $\sqrt{s}=13\,$\TeV\
during Run 2 of the Large Hadron Collider and correspond to an integrated luminosity of 139~\ifb .
A functional form is fitted to the dilepton invariant-mass distribution
to model the contribution from background processes, and a
generic signal shape is used to determine the significance of observed deviations from this
background estimate.
No significant deviation is observed and upper limits
are placed at the 95$\%$ confidence level on the fiducial cross-section times branching
ratio for various resonance width hypotheses. The derived limits are shown to be applicable to spin-0, spin-1 and spin-2 signal hypotheses.
For a set of benchmark models, the limits are converted into lower limits on the resonance mass and reach 4.5$\,$\TeV\ for the E$_6$-motivated $Z'_{\psi}$ boson.
Also presented are limits on Heavy Vector Triplet model couplings.
}
 
\author{The ATLAS Collaboration}
 
\AtlasRefCode{EXOT-2018-08}
 
\PreprintIdNumber{CERN-EP-2019-030}

\AtlasCoverSupportingNote{EXOT/LPX Dilepton Search full Run-2}{https://cds.cern.ch/record/2622420}
 
\AtlasCoverCommentsDeadline{8 April 2019}
 
\AtlasCoverAnalysisTeam{D. Abhayasinghe, T. Berry, M. Bugge, E. Dreyer, P. Falke, D. Hayden,
S. Hellesund, N. Hod, T. Hryn'ova , U. Klein, S. Lawlor, B. Li, P. Mastrandrea, V. Morisbak,
F. Ould-Saada, S. Rettie, S. Rosati, B. Stelzer, O. Stelzer-Chilton, D. Vannicola, A. White, B. Zhou}

\AtlasCoverEdBoardMember{Paul De Jong (chair), Giovanni Marchiori, Marco Vanadia}

\AtlasJournal{PLB}

\AtlasCoverEgroupEditors{atlas-EXOT-2018-08-editors@cern.ch}
 
\AtlasCoverEgroupEdBoard{atlas-EXOT-2018-08-editorial-board@cern.ch}
\hypersetup{pdftitle={ATLAS document},pdfauthor={The ATLAS Collaboration}}
 
\begin{document}
 
\maketitle

\section{Introduction}
\label{sec:intro}
Searches in the dilepton (dielectron and dimuon) final state have a long and illustrious history with
the discovery of the $J/\psi$ meson in 1974~\cite{PhysRevLett.33.1404, PhysRevLett.33.1406} and
$\Upsilon$ meson in 1977~\cite{1977PhRvL..39..252H} as well as the $Z$ boson in 1983~\cite{1983398,BAGNAIA1983130}. As these were
key steps which led to the establishment of the Standard Model
(SM) of particle physics, the study of the same final state could help to pave
the way to a better understanding of the physics processes beyond it.
 
Various models predict resonances which decay into dileptons and can be categorised
according to their spin. A new high-mass spin-0 resonance, $H$, introduced as part of an
extended scalar sector in some models, such as the Minimal Supersymmetric SM (MSSM)~\cite{Djouadi:2005gj},
has higher decay rate into a pair of muons rather than electrons. The majority of searches for new neutral
high-mass resonances have focused on a new spin-1 vector boson,
generally referred to as $Z'$, that appears in models with extended gauge symmetries.
Typical benchmark models include the Sequential Standard Model $Z'_\mathrm{SSM}$ boson~\cite{Altarelli:1989ff},
which has the same fermion couplings as the SM $Z$ boson, a $Z'_{\chi}$
and a $Z'_{\psi}$ boson of an E$_6$-motivated Grand Unification model~\cite{London:1986dk}, or
a $Z'_\mathrm{HVT}$ boson of the Heavy Vector Triplet model~\cite{delAguila:2010mx}.
In the first two models, the $Z'$ boson is a singlet, associated with a new U(1) gauge group, and generally
its couplings to the SM $W$ and $Z$ bosons are assumed to be zero.
The $Z'_\mathrm{HVT}$ boson is a neutral member of a new SU(2) gauge group, i.e.\ part of a triplet
and cannot exist without two new charged heavy bosons, $W'^{\pm}_\mathrm{HVT}$, with which it is nearly degenerate in mass.
New spin-2 resonances, excited states of the graviton, are introduced in
the Randall--Sundrum model~\cite{Randall:1999ee} with a warped extra dimension.
In experimental terms the described scenarios would result in a local excess of signal candidates over a smoothly falling dilepton mass spectrum.
This search has a clean experimental signature with a fully reconstructable final state and excellent detection efficiency.
 
This Letter presents a search for a new resonance decaying into two electrons or two muons
in  $139\,$\ifb{} of data collected in proton--proton ($pp$) collisions at the LHC
at a centre-of-mass energy $\sqrt{s}=13\,$\TeV. Previous searches with 36.1$\,$\ifb{} of $pp$ collision
data at $\sqrt{s}=13\,$\TeV\ conducted by the ATLAS and CMS experiments~\cite{EXOT-2016-05,CMS-EXO-16-047}
showed no significant excess and led to lower limits of up to 3.8$\,$\TeV\ for the mass of the $Z'_{\psi}$ boson.
The analysis presented in this Letter, compared with that published in Ref.~\cite{EXOT-2016-05}, benefits from:
a factor of four increase in integrated luminosity; several improvements in the reconstruction software, including
the use of a new dynamical, topological cell-clustering algorithm for electron reconstruction~\cite{ATL-PHYS-PUB-2017-022}
and an improved treatment of the relative alignment of the inner tracker and the muon tracking detectors in the muon reconstruction;
the use of invariant-mass sidebands of the expected signal in data to constrain the fit parameters of the background distribution, which is described by a smooth functional form instead of relying on simulation;
and a generic signal line shape described by a non-relativistic Breit--Wigner function convolved
with the detector resolution, which simplifies reinterpretations of the result.
 
\section{ATLAS detector}
\label{sec:models}
\newcommand{\AtlasCoordFootnote}{
ATLAS uses a right-handed coordinate system with its origin at the nominal interaction point (IP) in the centre
of the detector and the $z$-axis along the beam pipe. The $x$-axis points from the IP to the centre of the
LHC ring, and the $y$-axis points upwards. Cylindrical coordinates $(r,\phi)$ are used in the transverse plane,
$\phi$ being the azimuthal angle around the $z$-axis. The pseudorapidity is defined in terms of the polar angle
$\theta$ as $\eta = -\ln \tan(\theta/2)$. Angular distance is measured in units of $\Delta R \equiv \sqrt{(\Delta\eta)^{2} + (\Delta\phi)^{2}}$.}

ATLAS~\cite{PERF-2007-01,ATLAS-TDR-19,Abbott:2018ikt} is a multipurpose detector with a
forward--backward symmetric cylindrical geometry with respect to the LHC beam axis.\footnote{\AtlasCoordFootnote}
The innermost layers consist of tracking detectors in the pseudorapidity range
$|\eta|<2.5$. This inner detector (ID) is surrounded by a thin superconducting solenoid
that provides a \SI{2}{\tesla} axial magnetic field. It is enclosed by the electromagnetic and
hadronic calorimeters, which cover $|\eta|<4.9$. The outermost layers of ATLAS consist of an external
muon spectrometer (MS) within  $|\eta|<2.7$, incorporating three large toroidal magnetic assemblies with eight coils each.
The field integral of the toroids ranges between 2.0 and 6.0$\,$Tm for most of the acceptance.
The MS includes precision tracking chambers and fast detectors for triggering.
A two-level trigger system~\cite{TRIG-2016-01} reduces the recorded event rate to an average of \SI{1}{\kilo\hertz}.
 
\section{Data and simulation}
\label{sec:detector}
The dataset used in this analysis was collected during LHC Run 2 in stable beam conditions and with all detector
systems operating normally. The event quality was checked to remove events
with noise bursts or coherent noise in the calorimeters.
Events in the dielectron channel were recorded using a dielectron trigger based on the `very loose' or `loose'
identification criteria~\cite{TRIG-2016-01} with transverse energy (\et) thresholds
between 12 and 24$\,$\GeV\ for both electrons, depending on the data-taking period.
Events in the dimuon channel are required to pass at least one of two single-muon triggers:
the first requires a transverse momentum (\pt) of at least 50$\,$\GeV, while
the second has a threshold lowered to 26$\,$\GeV\ but requires the muon candidate to be isolated~\cite{TRIG-2016-01}.
The integrated luminosity of the dataset is determined to be $139.0\pm 2.4$~\ifb{}, following a methodology similar
to that detailed in Ref.~\cite{DAPR-2013-01},
and using the LUCID-2 detector for the baseline luminosity measurements~\cite{LUCID2},
from calibration of the luminosity scale using x-y beam-separation scans.
 
While the search in this analysis is carried out entirely in a data-driven way,
simulated event samples for the signal and background processes are used to
determine appropriate functions to fit the data, study background compositions and to evaluate the signal efficiency.
The main backgrounds in decreasing order of importance are Drell--Yan (DY),
top-quark pair ($t\bar{t}$), single-top-quark
and diboson production.
Multi-jet and $W+$jets processes in the dielectron channel
are estimated with a data-driven method~\cite{EXOT-2016-05}.
Multi-jet and $W+$jets processes in the dimuon channel as well as processes with
$\tau$-leptons in both channels have a negligible impact and are not considered.
The Monte Carlo (MC) event generators for the hard-scatter process,
showering and
parton distribution functions (PDFs) are listed in Table~\ref{tab:MC}.
The `afterburner' generators such as  \textsc{Photos}~\cite{Golonka:2005pn} for
the final-state photon radiation (FSR) modelling,
\textsc{MadSpin}~\cite{Artoisenet:2012st} to preserve top-quark spin correlations, and
\textsc{EvtGen}~\cite{Lange:2001uf}, used for the modelling of
$c$- and $b$-hadron decays, are also reported.
 
\begin{table}[htbp]
\centering
\caption{The event generators used for simulation of the signal and background processes.
The acronyms ME and PS stand for matrix element and parton shower.
The top-quark mass is set to 172.5$\,$\GeV.
}
\scriptsize
\begin{tabular}{lll}
\hline
Background Process & ME Generator and ME PDFs & PS and non-perturbative effect with PDFs \\\hline
NLO Drell--Yan & \POWHEGBOX~\cite{Alioli:2010xd,Frixione:2007vw}, CT10~\cite{ct10}, \textsc{Photos} & \PYTHIAV{v8.186}~\cite{pythia8}, CTEQ6L1~\cite{ATL-PHYS-PUB-2014-021,Stump:2003yu}, \textsc{EvtGen1.2.0} \\
$t\bar{t}$  & \POWHEGBOX, NNPDF3.0NLO~\cite{Ball:2014uwa} & \PYTHIAV{v8.230}, NNPDF23LO~\cite{Ball:2012cx}, \textsc{EvtGen1.6.0} \\
Single top $s$-channel, $Wt$& \POWHEGBOX, NNPDF3.0NLO & \PYTHIAV{v8.230}, NNPDF23LO, \textsc{EvtGen1.6.0} \\
Single top $t$-channel & \POWHEGBOX, NNPDF3.04fNLO, \textsc{MadSpin} & \PYTHIAV{v8.230}, NNPDF23LO, \textsc{EvtGen1.6.0}  \\
Diboson ($WW$, $WZ$ and $ZZ$) & \SHERPA 2.1.1~\cite{Gleisberg:2008ta}, CT10 &\SHERPA 2.1.1, CT10  \\\hline
Signal Process & & \\\hline
LO Drell--Yan & \PYTHIAV{v8.186}, NNPDF23LO  &  \PYTHIAV{v8.186}, NNPDF23LO, \textsc{EvtGen1.2.0} \\
Randall--Sundrum $G^*\rightarrow\ell\ell$ & \PYTHIAV{v8.210}, NNPDF23LO  & \PYTHIAV{v8.210}, NNPDF23LO, \textsc{EvtGen1.2.0} \\
MSSM $gg\rightarrow H\rightarrow\ell\ell$ & \POWHEGBOX, CT10 & \PYTHIAV{v8.212}, CTEQ6L1, \textsc{EvtGen1.2.0} \\\hline
\end{tabular}
\normalsize
\label{tab:MC}
\end{table}
 
The DY~\cite{ATL-PHYS-PUB-2016-003} and diboson~\cite{ATL-PHYS-PUB-2016-002} samples are generated in slices of
dilepton mass to increase the sample size in the high-mass region.
Next-to-next-to-leading-order (NNLO) corrections in quantum chromodynamic (QCD) theory and next-to-leading-order (NLO) corrections 
in electroweak (EW) theory, are calculated and applied to the
DY events. 
The corrections are computed with {\textsc{VRAP}} v0.9~\cite{vrap}
and the CT14 NNLO PDF set~\cite{CT14} in the case of QCD effects
whereas they are computed with {\textsc{MCSANC}}~\cite{MCSANC} in the case of quantum electrodynamic effects
due to initial state radiation, interference between initial and final state radiation, and
Sudakov logarithm single-loop corrections.
The top-quark samples~\cite{ATL-PHYS-PUB-2016-020} are normalised to the
cross-sections calculated at NNLO in QCD including
resummation of the next-to-next-to-leading logarithmic soft gluon terms as provided by
\textsc{Top++}2.0~\cite{Czakon:2011xx}.
 
Spin-1 signal templates are obtained by a matrix-element reweighting~\cite{EXOT-2016-05} of the leading-order (LO)
DY samples generated in slices of dilepton mass. These signal templates are used only for
cross-section and efficiency calculations. The relative natural width ($\Gamma_{Z'}/m_{Z'}$)
for the benchmark models considered varies between 0.5$\%$ for $Z'_{\psi}$ and 3$\%$ for $Z'_\mathrm{SSM}$.
Interference effects between the resonant signal and the background processes are neglected.
Higher-order QCD corrections for all the spin-1 signals are computed with the same methodology as for
the DY background. For the HVT model, these corrections are not applied,
which ensures consistent treatment with the other signal channels in an eventual combination, similar to that described in Ref.~\cite{EXOT-2017-31}.
Electroweak corrections are not applied to the signal samples due to their large model dependence.
Spin-0 signal efficiencies are obtained from samples of the MSSM gluon--gluon fusion production of a heavy Higgs boson decaying into dilepton pairs,
$gg\rightarrow H\rightarrow\ell\ell$, produced in the mass range $m_H=400$--$1000\,$\GeV\ and with relative
natural width ($\Gamma_H/m_H$) varying between zero and 20$\%$.
Spin-2 signal efficiencies are obtained from Randall--Sundrum graviton $G^*\rightarrow\ell\ell$ samples produced in the
mass range $m_{G^*}=750$--$5000\,$\GeV\ and with coupling strengths, $k/\overline{m}_\mathrm{Pl}$, of 0.1, 0.2 and 0.3,
where $k$ is a scale that defines the warp factor of the extra dimension and $\overline{m}_\mathrm{Pl}$ is the reduced Planck mass.
 
Simulated event samples include the effect of multiple $pp$ interactions in the same
or neighbouring bunch crossings. These effects are collectively referred to as pile-up.
The simulation of pile-up collisions was performed with
\PYTHIAV{v8.186} using the ATLAS A3 set of tuned parameters~\cite{ATL-PHYS-PUB-2016-017} and the NNPDF23LO PDF set,
and weighted to reproduce the average number of pile-up interactions per bunch crossing observed in data.
The generated events were passed through a full detector simulation~\cite{SOFT-2010-01} based on\ \GEANT~4~\cite{geant}.
Spin-0 and spin-2 MC signal samples were produced with a fast parameterisation of the calorimeter response~\cite{ATL-PHYS-PUB-2010-013}.
 
Very large generator-level-only MC samples (with more than 55 times the data events) for NLO DY events
are used for the background studies described in Section~\ref{sec:signal}.
These samples could not be produced with the full detector simulation due to the large number of events required.
 
\section{Event selection}
\label{sec:analysis}
\newcommand{\ConeSize}{$\Delta R$ has a maximum value of 0.3 and decreases as a function of $\pt$ as 10\,GeV/\pt{}[\GeV].
}
 
The selection of dilepton events closely follows that described in Ref.~\cite{EXOT-2016-05}.
An event is selected if at least one $pp$ interaction
vertex is reconstructed. The primary vertex is chosen to be the vertex with
the highest summed $\pt^2$ of tracks with transverse momentum $\pt>0.5\,$\gev{} which are associated with the vertex.
 
Electron candidates are reconstructed from ID tracks that are matched to clusters of energy deposited in
the electromagnetic calorimeter
with energy deposition consistent with that of an electromagnetic shower~\cite{PERF-2017-01}.
Reconstructed electrons must have \et$>30\,$\gev{}, satisfy $|\eta|<2.47$ in order
to pass through the fine-granularity region of the EM calorimeter, and
be outside the range $1.37 < |\eta| < 1.52$ corresponding to the transition
region between the barrel and endcap EM calorimeters.
The calorimeter granularity in the excluded transition region is reduced,
and the presence of significant additional inactive material degrades the
electron
identification capabilities and energy resolution.
The `medium' electron working point used for the final selection
has an identification and reconstruction efficiency for prompt electrons above 92$\%$ for \et$>80\,$\GeV.
 
Muon candidates are identified by matching ID tracks to tracks reconstructed in the MS~\cite{PERF-2015-10}.
Muon candidates must have \pt$>30\,$\gev{} and $|\eta|<2.5$.
To ensure optimal muon momentum resolution at high \pt,
the `high \pt' identification working point is used. It requires at least three hits in each
of three layers of precision tracking chambers in the MS, and specific regions of the MS where
the alignment is suboptimal are vetoed as
a precaution.
These requirements reject about 80\% (13\%) of the muon candidates
in (outside) the barrel--endcap overlap region, $1.01< |\eta|<1.1$.
The muon `high \pt' working point has an $\eta$-averaged
efficiency of 69$\%$ at 1$\,$\TeV\ which decreases to 64$\%$ at 2.5$\,$\TeV\
due to increased occasional catastrophic energy loss at high \pt.
Additionally, a `good muon' selection requires that the uncertainty in the charge-to-momentum ratio
of muon candidates is less than a
\pt-dependent value. This selection is fully efficient below 1$\,$\TeV,
but introduces an additional inefficiency of 7$\%$ at 2.5$\,$\TeV.
 
Electron (muon) candidate tracks must be consistent with the primary vertex both along the beamline,
where the longitudinal impact parameter $z_0$ is required to satisfy $|z_0 \ \mathrm{sin}\theta|<0.5\,$mm,
and in the transverse plane, where the transverse impact parameter significance $|d_0/\sigma(d_0)|$
is required to be less than 5 (3).
To reduce background from misidentified jets as well as from light- and heavy-flavour
hadron decays inside jets, lepton candidates are required to be isolated. Electrons must pass the `gradient' isolation working point which
targets an \et-dependent value of the isolation efficiency, uniform in $\eta$, using
a combination of track and calorimeter isolation requirements~\cite{PERF-2017-01}.
For muons, the summed scalar \pt of good-quality tracks with
\pt$>1\,$\GeV\ originating from the primary vertex within a cone of variable size\footnote{\ConeSize} $\Delta R$
around the muon, but excluding the
muon-candidate track itself, must be less than 6$\%$ of the \pt{} of the muon candidate.
The efficiency of this selection is above 99$\%$ for both electrons and muons with \pt$>60\,$\GeV.
Corrections are applied to electron (muon) candidates to match the energy (momentum) scale and resolution between simulation and data. These corrections are derived in an energy independent way for electrons~\cite{PERF-2017-03}.
For muons, the correction is determined as a function of \pT{} up to 300$\,$\GeV{}, from a fit to $Z\rightarrow\mu\mu$ data with templates derived from simulation~\cite{PERF-2015-10}. At high transverse momentum, the calibrations are dominated by corrections extracted from alignment studies, using special runs with the toroidal magnetic field off.
Corrections to the lepton efficiencies in the simulation are derived from the data for electron \et{} (muon \pT) up to 150 (200)$\,$\GeV~\cite{PERF-2017-01,PERF-2015-10}. The simulation is used to extrapolate to higher electron \et{} (muon \pt) and to study systematic effects.
 
The events are required to contain at least two same-flavour leptons.  If additional leptons are present in the event,
the two same-flavour leptons with the largest \et{}~(\pt) in the electron (muon) channel are selected to form the dilepton pair.
If two different-flavour pairs are found, the dielectron pair is kept,
because of the better resolution and higher efficiency for electrons.
A selected muon pair is required to be oppositely charged.
For an electron pair, the opposite-charge requirement is not applied because of the higher probability
of charge misidentification for high-\et{} electrons. The reconstructed mass of the dilepton system
after the full analysis selection, $m_{\ell\ell}$, is required
to be above 225$\,$\GeV\ to avoid the $Z$ boson peak region, which cannot be
described by the same parameterisation as the high-mass part of the dilepton
distributions.

\section{Reconstructed dilepton mass modelling}
\label{sec:mass}
 
The relative dilepton mass resolution is defined as
$(m_{\ell\ell}-m_{\ell\ell}^\mathrm{true})/m_{\ell\ell}^\mathrm{true}$, where $m_{\ell\ell}^\mathrm{true}$
is the generated dilepton mass at Born level before FSR.
The mass resolution is parameterised as a sum of a Gaussian distribution,
which describes the detector response, and a Crystal Ball function
composed of a secondary Gaussian distribution with a power-law low-mass tail, which accounts for bremsstrahlung
effects in the dielectron channel or for the effect of poorly reconstructed muons.
The parameterisation of the relative dilepton mass resolution as a function
of $m_{\ell\ell}^\mathrm{true}$ is determined by a simultaneous fit of the function described above
to NLO DY MC events. The MC sample is separated in
200 $m_{\ell\ell}^\mathrm{true}$ bins of equal size on a logarithmic scale in the range of 130$\,$\GeV\ to 6$\,$\TeV.
This procedure is repeated to evaluate the uncertainty on the fit parameters
by shifting individually the lepton energy and momentum scale and resolutions
by their uncertainties.
 
\section{Signal and background modelling}
\label{sec:signal}
A resonant signal is searched for by fitting the data dilepton mass distribution.
The fit function consists of a smooth functional form for the background,
and a generic signal shape.
The generic signal shapes are constructed from non-relativistic Breit--Wigner functions of
various widths convolved with the detector resolution, obtained as described in the previous section.
The shape of the dilepton invariant mass distribution
for a signal resonance with intrinsic width that is negligible
compared with the detector resolution
(zero-width signal) is obtained from the mass resolution only.
 
To allow for a generic resonance search, a fiducial region at particle level is defined
following the selection criteria applied to the reconstructed lepton candidates:
each electron and muon candidate needs to pass $|\eta|<2.5$ and \et(\pt)$\,>30\,$\gev{},
and the dilepton mass has to satisfy $m_{\ell\ell}^\mathrm{true}>m_{X}-2 \Gamma_{X}$, where $m_{X}$ and
$\Gamma_{X}$ represent the pole mass and width of a hypothetical resonance $X$,
respectively. This selection is added in order to reduce the model dependence from off-shell effects.
 
The nominal combined reconstruction and identification efficiency in the fiducial region
is extracted from the DY sample and thus assumes the kinematics of a spin-1 boson.
For the dielectron (dimuon) channels, it
varies from 64\% (54\%) at 225$\,$\gev{} to 74\% (38\%) at 6$\,$\tev{} for the zero-width signals.
For a spin-1 signal with 10$\%$ relative width, the efficiency changes by less than $0.5\%$ relative to
a signal with zero width for both channels over most of the considered invariant-mass range.
Only above 5$\,$\tev{} in the dimuon channel are the variations as large as 2$\%$ in absolute efficiency.
For the spin-0 and spin-2 samples, width-related variations are below 1$\%$.
For the dielectron channel, spin-0 and spin-2 efficiencies are higher
than the corresponding spin-1 values by at most 4$\%$.
For the dimuon channel, efficiencies for spin-0 and spin-2 signals
are at most 1$\%$ lower than the corresponding spin-1 values.
The systematic uncertainties of the overall efficiency are due to the
uncertainties in the trigger,
isolation, identification, and reconstruction efficiencies.
 
The smooth functional form for the background is based on fit performance studies on a MC background template.
The associated uncertainties are also estimated through
these studies. In order to minimise the statistical uncertainties in this procedure,
the background template for DY is produced from large-statistics
samples simulated only at generator level and smeared by the experimental dilepton mass resolution, described in the previous section,
with mass-dependent acceptance and efficiency corrections applied.
A similar procedure is applied to the generator-level dilepton mass distribution in the
$t\bar{t}$ sample exploiting the larger number of events from the generator-level mass distribution.
The distributions from the diboson and single-top simulated samples
and, in the electron channel, a template for multi-jet and $W+$jet processes are also considered.
All MC-based contributions are scaled by their respective cross-sections and
summed together to obtain the background template for the choice of the smooth functional form.
 
In order to select the background functional form,
a fit to the dilepton mass background template is performed,
under the signal plus background hypothesis,
for various functional forms, following the procedure outlined in Ref.~\cite{HIGG-2013-08}.
The chosen functional form is the one with the smallest absolute number
of fitted signal events (`spurious signal'), which are determined as as a function of $m_{\ell\ell}$:
 
\begin{align}
\label{eq:GlobalFit_ll}
f_{\ell\ell}(m_{\ell\ell}) &= f_{\mathrm{BW},Z}(m_{\ell\ell}) \cdot \left(1 - x^{c}\right)^{b} \cdot x^{\sum_{i=0}^3 p_i\log(x)^i},
\end{align}
 
where $x = m_{\ell\ell}/\sqrt{s}$ and parameters
$b$ and $p_i$ with $i=0,..3$ are left free in the fit to data and independent for dielectron and dimuon channels.
The parameter $c$ is $1$ for the dielectron and $1/3$ for the dimuon channel.
The function $f_{\mathrm{BW},Z}(m_{\ell\ell})$ is a non-relativistic Breit--Wigner function with $m_Z = 91.1876\,\GeV$ and $\Gamma_Z = 2.4952\,\GeV$~\cite{PhysRevD.98.030001}.
The normalisation of the background function is such that the
integral $a$ corresponds to the total number of background events.
To further validate this functional form an extra degree of freedom ($i=4$) is added
to the fit function before the final data analysis, to
check if it improves the likelihood value of the fit by more than 2$\sigma$.
To check the fit stability in the high-mass region, signal injection tests
are performed at various mass points.
No significant bias in the number of extracted signal events is observed.
 
Uncertainties related to the background modelling are
propagated into the determination of the spurious signal. Smooth templates for systematic
shape uncertainties are produced using the same procedure as for the nominal templates.
The uncertainties considered include variations due to PDFs~\cite{EXOT-2016-05} and
normalisation of the $t\bar{t}$ background component~\cite{TOPQ-2015-09}. Uncertainties on the
multi-jet and $W+$jet background contributions~\cite{EXOT-2016-05} are also considered in the dielectron channel.
For the selected function, the largest spurious signal (accounting for all systematic variations)
is required to be less than 30$\%$ of the
statistical uncertainty in the fitted signal yield
(from the background distribution) for the zero-width signal. This criterion is
relaxed to 50$\%$ for signals of greater width.
The systematic uncertainty of the background estimate is mass dependent and corresponds to a functional
interpolation between the highest maxima among the spurious-signal-yield distributions for all systematic variations.
The spurious-signal yield is calculated independently for the relative signal width assumptions
between zero and 10$\%$ in steps of $0.5\%$.
 
The impact of systematic uncertainties on the signal yield is shown in Table~\ref{tab:syst}.
Only systematic uncertainties which change the fitted signal yield by more than 0.5$\%$
at any point in the mass spectrum are considered.
The largest systematic uncertainty at low mass in both channels originates from the spurious signals.
The largest systematic uncertainty in the dielectron channel at high mass originates from the
electron identification efficiency.
The uncertainty associated with the `good muon' requirement is dominant in the dimuon channel
at high mass. This uncertainty is estimated with a conservative approach in a dataset collected in 2015--2016,
corresponding to 36$\,$fb$^{-1}$, by comparing efficiencies obtained in data and in simulation.
 
\begin{table}[htbp]
\centering
\caption{
The relative impact of $\pm 1\sigma$ variation of systematic uncertainties on the signal yield in percent
for zero (10$\%$) relative width signals at the pole masses of $300\,\GeV$, 2 and $5\,\TeV$
for dielectron and dimuon channels. Sources of uncertainties leading to an impact smaller than
0.5\% on the signal yield at any point of the mass mass spectrum are not shown.
A signal is injected at the cross-section limit.
}
\vspace{0.3cm}
\begin{tabular}{l|cccccc}
\hline
Uncertainty source &  \multicolumn{3}{c}{Dielectron} & \multicolumn{3}{c}{Dimuon}  \\
for $m_X$ [\GeV] &  300 & 2000 & 5000 & 300 & 2000 & 5000  \\ \hline
Spurious signal & $\pm$12.5 (12.0) & $\pm$4.6 (10.8) & $\pm$0.1 (1.0) & $\pm$11.7 (11.0) & $\pm$3.8 (3.5) & $\pm$2.1 (2.2) \\
\\[-1em]
Lepton identification  & $\pm$1.6 (1.6) & $\pm$5.6 (5.6) & $\pm$5.6 (5.6) &  $\pm$1.8 (1.8) & $_{-10}^{+12}$  \Big($_{-10}^{+12}$\Big) & $_{-20}^{+25}$  \Big($_{-20}^{+25}$\Big) \\
\\[-1em]
Isolation & $\pm$0.3 (0.3) & $\pm$1.1 (1.2) & $\pm$1.1 (1.1) &$\pm$0.4 (0.4) &$\pm$0.4 (0.4) &$\pm$0.4 (0.5)  \\
\\[-0.5em]
Luminosity & $\pm$1.7 (1.7) & $\pm$1.7 (1.7) & $\pm$1.7 (1.7) & $\pm$1.7 (1.7) & $\pm$1.7 (1.7) & $\pm$1.7 (1.7) \\
\\[-1em]
Electron energy scale & $_{-4.0}^{-1.7}$ \Big($_{-1.8}^{+1.0}$\Big) & $_{-6.0}^{-1.9}$ \Big($_{-2.9}^{+1.7}$\Big) & $^{+0.1}_{-0.4}$ ($\pm$0.8) & - & - & - \\
\\[-1em]
Electron energy resolution & $_{-8.3}^{+7.9}$ \Big($_{-0.9}^{+1.1}$\Big) & $_{-11.8}^{+9.0}$ \Big($_{-0.5}^{+0.7}$\Big) & $_{-0.9}^{+0.4}$ ($\pm$0.1) & - & -  & - \\
\\[-1em]
Muon ID resolution & - & - & - & $_{-2.3}^{+0.8}$ \Big($_{-0.8}^{+0.3}$\Big) & $_{-1.3}^{+0.9}$ \Big($_{-1.1}^{+0.7}$\Big) & $_{-0.4}^{+0.6}$ \Big($_{-0.3}^{+0.5}$\Big) \\
\\[-1em]
\\[-1em]
Muon MS resolution & - & - & - &  $_{-3.8}^{+2.8}$ \Big($_{-1.3}^{+1.0}$\Big) &  $_{-3.0}^{+3.2}$ \Big($_{-2.4}^{+2.6}$\Big) & $\pm$2.4 (2.1) \\
\\[-1em]
`Good muon' requirement  & - & - & - & $\pm$0.6 (0.6) &   $_{-8.2}^{+9.0}$ \Big($_{-8.2}^{+9.0}$\Big) &   $_{-35}^{+55}$ \Big($_{-35}^{+55}$\Big) \\\hline
\end{tabular}
\normalsize
\label{tab:syst}
\end{table}
 
\section{Statistical analysis}
\label{sec:statistics}
The numbers of signal and background events, as a function of the signal mass and width hypothesis, are
estimated from simultaneous maximum-likelihood fits of the signal-plus-background models
to the data $m_{\ell\ell}$ distribution.
Systematic uncertainties are included in the fits via nuisance parameters constrained by
penalty terms which are either Gaussian (e.g. energy and momentum scale uncertainties) or log-normal (efficiency
and resolution uncertainties).
Potential mismodelling of the background estimate is accounted for through an additional nuisance parameter
allowing non-zero signal normalisation under the null hypothesis constrained by the measured spurious signal.
Dielectron and dimuon channels are considered
both as independent channels and in a combined approach, under a lepton-flavour universality assumption~\cite{Altarelli:1989ff,London:1986dk}.
 
The significance of a signal is summarised by a $p$-value, the probability of observing an excess at least as signal-like as the one observed in data, in the absence of signal.
The local $p$-value of the background-only hypothesis ($p_0$)
is determined from a profile-likelihood-ratio-test statistic~\cite{Cowan:2010js}
as detailed in Ref.~\cite{HIGG-2016-08} in the asymptotic approximation.
Global significance values are also computed in the asymptotic approximation to account for the
trial factors due to scanning the signal mass hypothesis~\cite{Gross:2010qma}.
Upper limits at the 95$\%$ confidence level (CL)
are set on the fiducial cross-section times branching ratio into the corresponding
dilepton final state, given the integrated luminosity of the data and the signal efficiency.
The limits are evaluated with the modified frequentist CL$_\mathrm{S}$ method~\cite{Read:2002hq}
using the asymptotic approximation to the test-statistic distribution~\cite{Cowan:2010js}. Cross-checks
with sampling distributions generated using pseudo-experiments are used to test the accuracy of this approximation
for the high-mass part of the dilepton spectra. The approximation is found to lead to limits that are
stronger than those obtained with pseudo-experiments above $3\,\TeV$. This effect
reaches 25\% (35\%) at $5\,\TeV$ ($6\,\TeV$) for the combined dilepton channel. The impact of this approximation on the mass limits is below $100\,\GeV$.
 
\section{Results}
\label{sec:result}
\newcommand{\params}{{
The resulting fit parameters for dielectron channel are:
$a = 178000 \pm 400$,
$b = 1.5 \pm 1.0$,
$p_0 = -12.38 \pm 0.09$,
$p_1 = -4.295 \pm 0.014$,
$p_2 = -0.9191 \pm 0.0027$,
$p_3 = -0.0845 \pm 0.0005$; for dimuon channel are:
$a = 138700 \pm 400$,
$b = 11.8 \pm 0.5$,
$p_0 = -7.38 \pm 0.12$,
$p_1 = -4.132 \pm 0.017$,
$p_2 = -1.0637 \pm 0.0029$,
$p_3 = -0.1022 \pm 0.0005$.}}
 
The dilepton invariant-mass distributions for the events that pass the full analysis
selection are shown in Figure~\ref{fig:fits}.
The event with highest reconstructed mass
is a dielectron candidate with $m_{ee} = 4.06\,\TeV$, formed of two electrons with $E_\mathrm{T} = 2.01\,\TeV$ and $E_\mathrm{T} = 1.92\,\TeV$ in the barrel region of the calorimeter. The event with highest reconstructed mass
in the dimuon channel has an invariant mass of $m_{\mu\mu} = 2.75\,\TeV$. Both muon candidates are in the barrel section of the muon spectrometer and their transverse momenta are
$p_\mathrm{T} = 1.82\,\TeV$ and $p_\mathrm{T} = 1.04\,\TeV$.
 
The fit to data\footnote{\params} is performed in bins of $1\,\GeV$ and uses the function in Eq.~(\ref{eq:GlobalFit_ll}).
In both channels, validation tests using the extension of the functional form described in Section~\ref{sec:signal}
did not yield any significant improvement,
so the function in Eq.~(\ref{eq:GlobalFit_ll}) is used without modification.
 
\begin{figure}[!htpb]
\centering
\subfloat[][]{      {\includegraphics[width=0.45\textwidth]{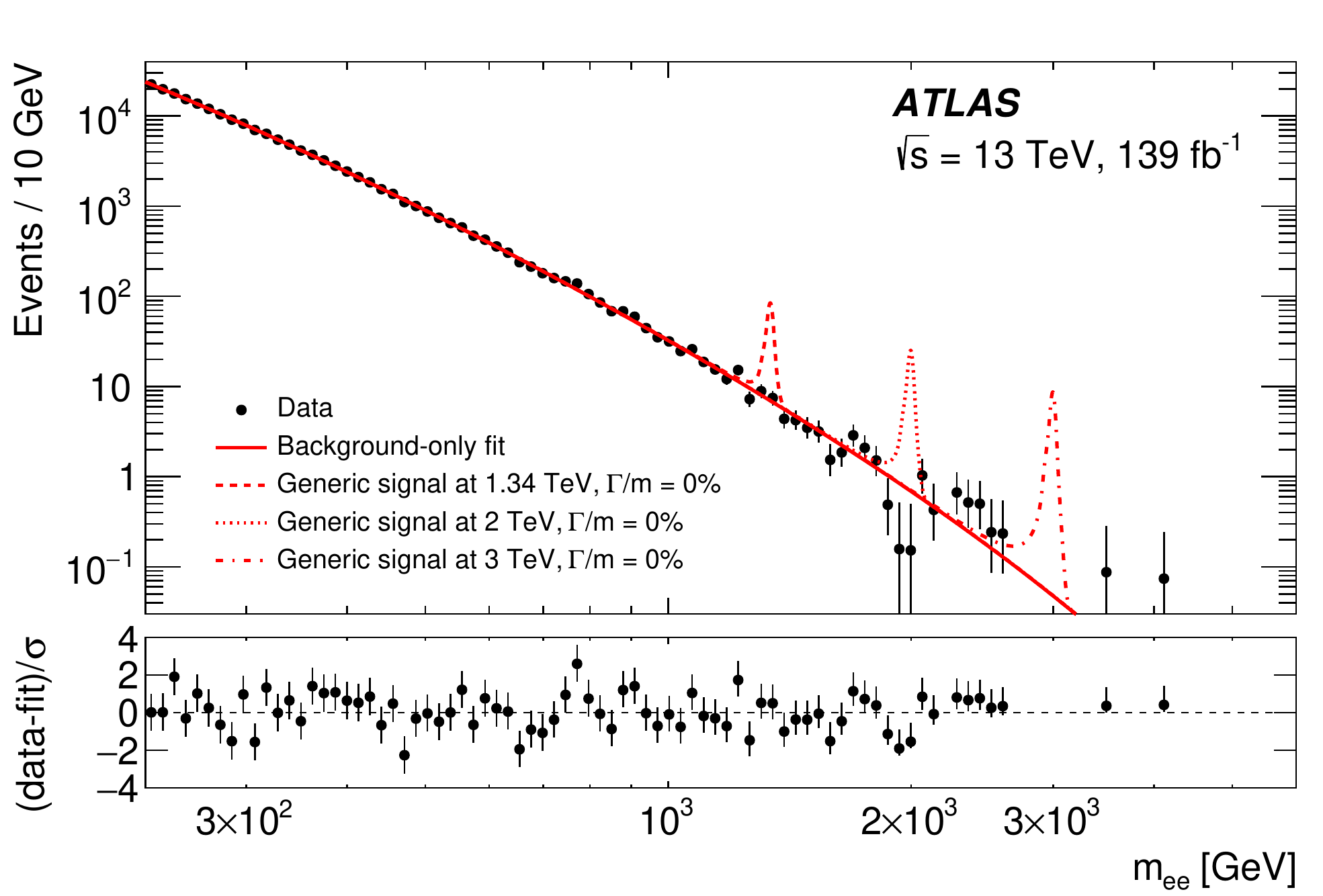}}}
\subfloat[][]{     {\includegraphics[width=0.45\textwidth]{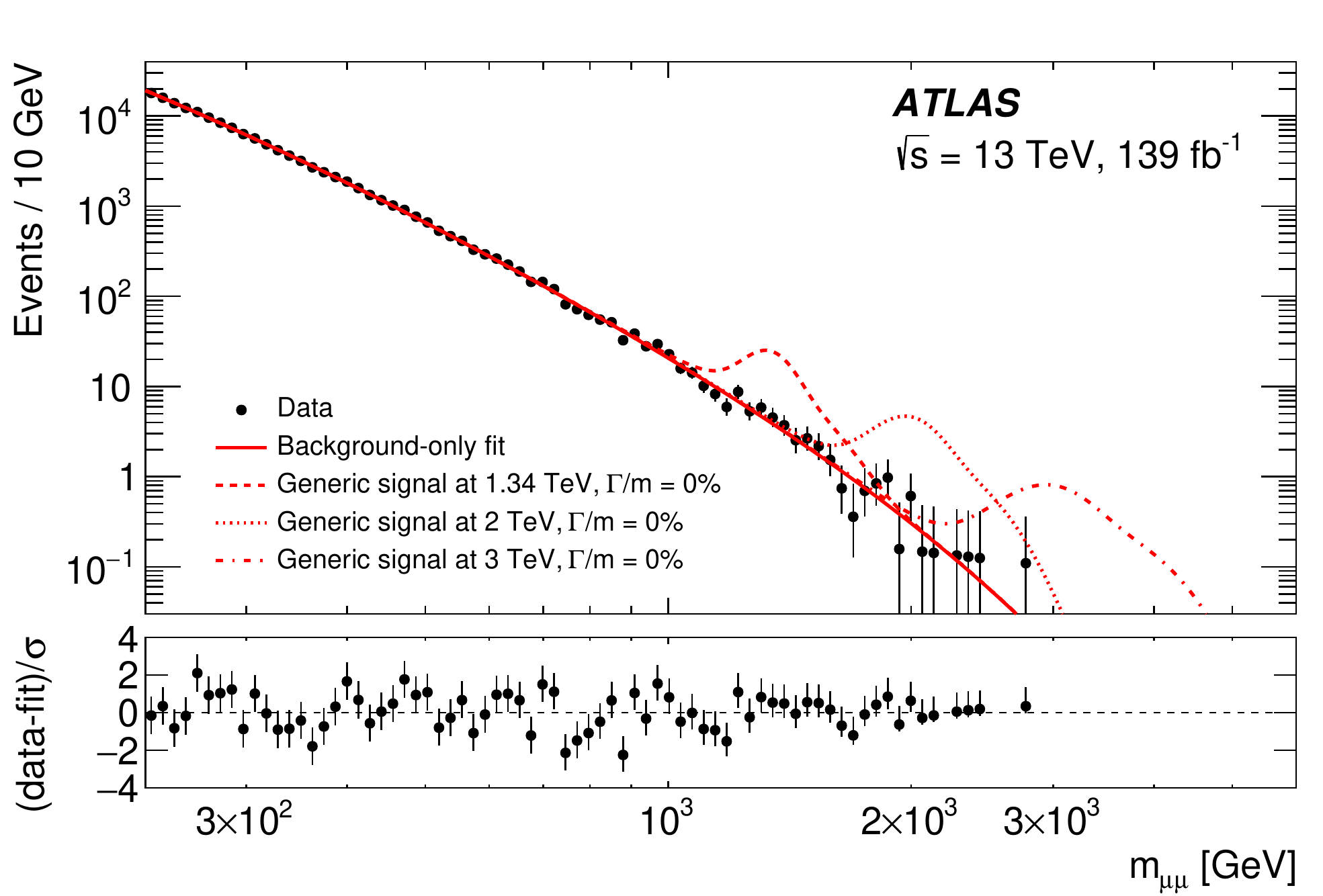}}}
\caption{
Distribution of the (a) dielectron and (b) dimuon invariant mass for events passing
the full selection. Generic zero-width signal shapes, scaled to 20 times the value of the
corresponding expected upper limit at 95\% CL on the fiducial cross-section times branching ratio,
with pole masses of $m_X=$1.34, 2 and $3\,\TeV$
as well as background-only fits are superimposed. The data points are plotted
at the centre of each bin. The error bars indicate statistical uncertainties only. The differences
between the data and the fit results in units of standard deviations of the statistical uncertainty are shown in the bottom panels.
}
\label{fig:fits}
\end{figure}
 
\begin{figure}[!htpb]
\centering
{\includegraphics[width=0.45\textwidth]{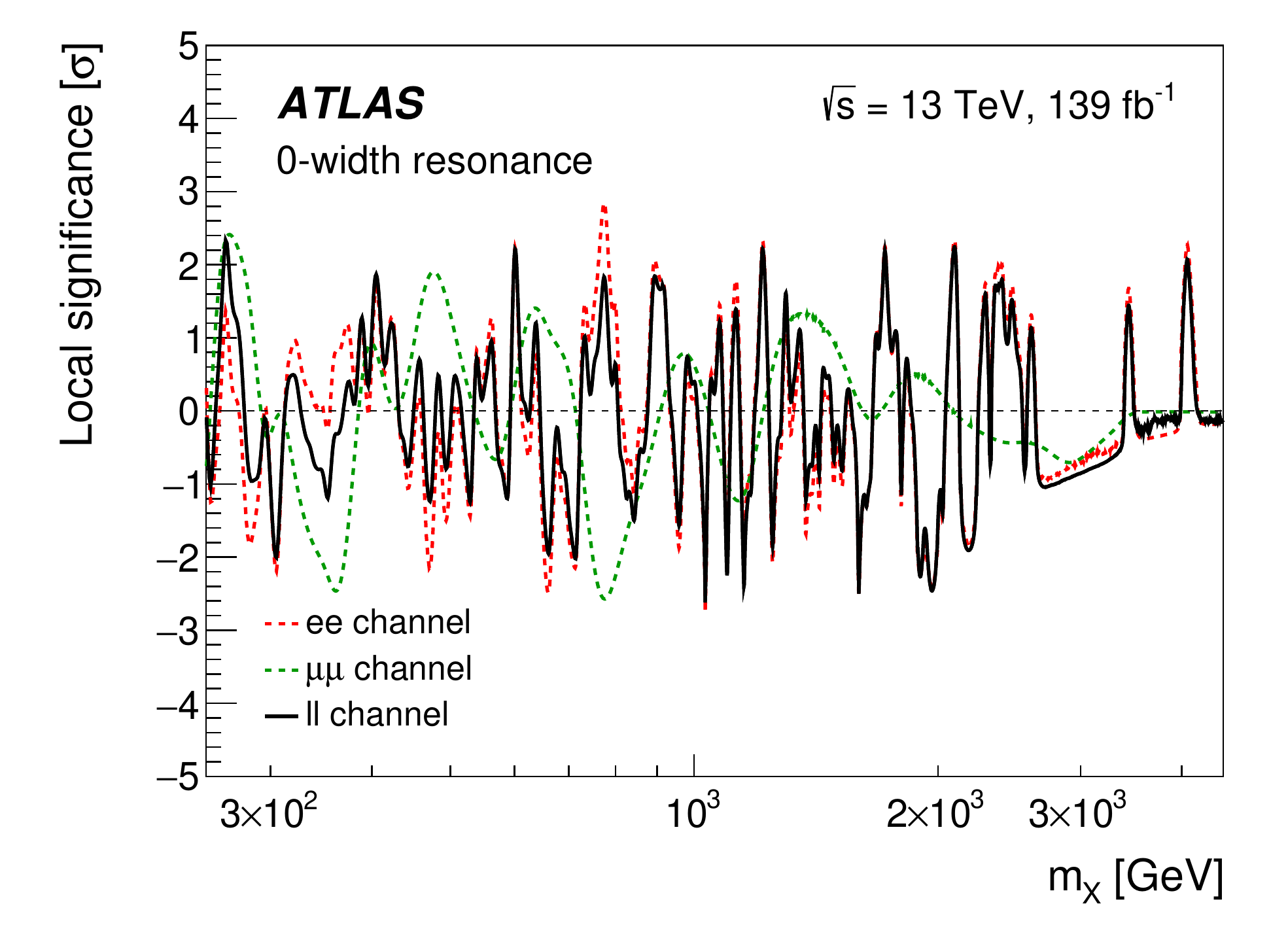}}
\caption{Probability that the observed spectrum is compatible with the background-only hypothesis for the dielectron, dimuon and combined dilepton channels.
The local $p_0$ is quantified in standard deviations $\sigma$ as a function of pole mass $m_{X}$.
}
\label{fig:scan}
\end{figure}
 
\begin{figure}[!htpb]
\centering
\subfloat[][]{{\includegraphics[width=0.485\textwidth]{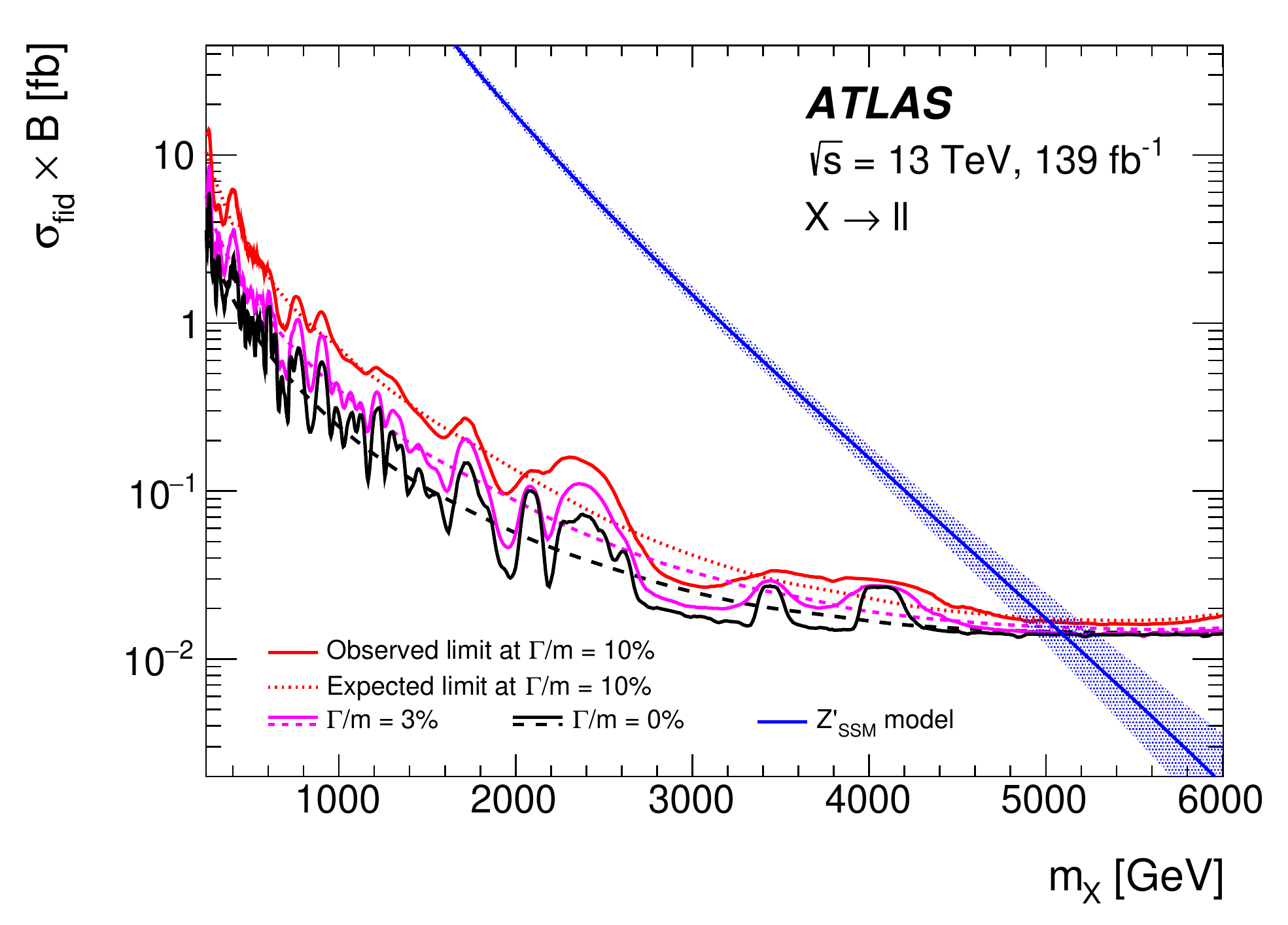}}}
\subfloat[][]{{\includegraphics[width=0.485\textwidth]{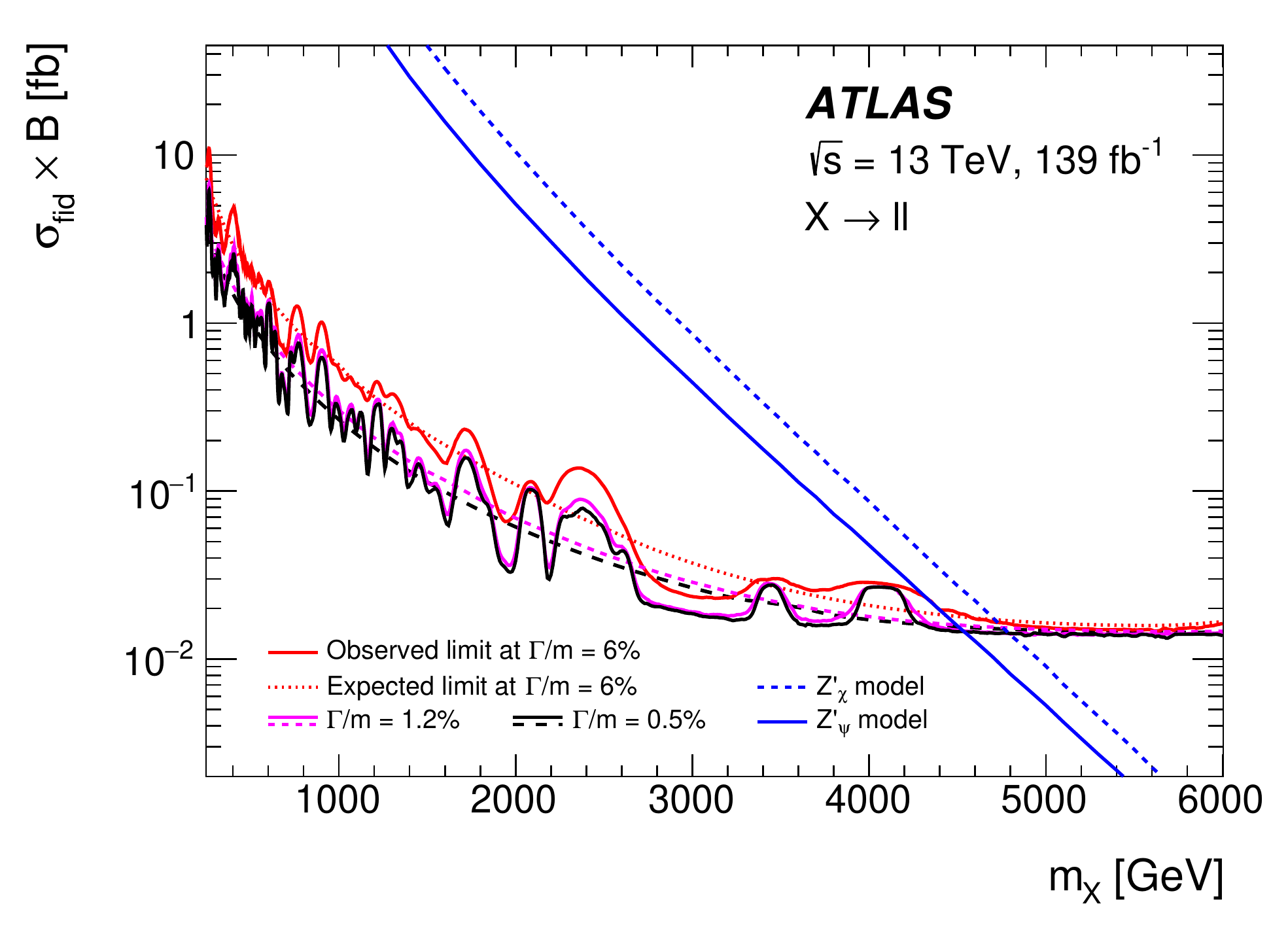}}}
\caption{
Upper limits at 95$\%$ CL on the fiducial cross-section times branching ratio
as a function of pole mass for (a) the zero-width, 3$\%$, 10$\%$
and (b) 0.5$\%$, 1.2$\%$, 6$\%$ relative width signals for the combined dilepton channel.
Observed limits are shown as a solid line and expected limits as a dotted/dashed line.
Also shown are theoretical cross-sections for (a) $Z'_\mathrm{SSM}$ ($\Gamma/m = 3.0\%$) and (b) $Z'_{\chi}$ ($\Gamma/m = 1.2\%$) and $Z'_{\psi}$ ($\Gamma/m = 0.5\%$) in the fiducial region.
The signal theoretical uncertainties are shown as a band on the $Z'_\mathrm{SSM}$ theory line
and are derived as in Ref.~\cite{EXOT-2016-05}. They are shown for illustration purposes, but are not included in the limit calculation.
}
\label{fig:limit}
\end{figure}

The probability that the data are compatible with the background-only hypothesis is shown in Figure~\ref{fig:scan} as a
function of pole mass for zero-width signals. No significant excess is observed.
The largest deviations from the background-only hypothesis in the dielectron, dimuon and combined dilepton channels are
observed at masses of $774\,\GeV$, $267\,\GeV$ and $264\,\GeV$ for zero-width signals with a local $p_0$ of 2.9$\,\sigma$, 2.4$\,\sigma$ and 2.3$\,\sigma$
and a global significance of $0.1\,\sigma$, $0.3\,\sigma$, and zero, respectively.
 
Figure~\ref{fig:limit} shows the upper limits on the fiducial cross-section
times branching ratio to two leptons of a single flavour for generic resonances of various relative widths as a function of their mass.
The observed limits for pole masses ranging from 250 to $750\,\GeV$ are obtained with a spacing of $1\,\GeV$. The granularity is reduced above that mass,
but remains below the experimental resolution of the $ee$ channel.
The observed limit on the fiducial cross-section times branching ratio ranges from 3.6 (13.1)$\,$fb
at $250\,\GeV$ to about 0.014 (0.018)$\,$fb at $6\,\TeV$ for the zero (10$\%$) relative width signal
in the combined dilepton channel.
The impact of systematic uncertainties on this search is small across all mass and width assumptions,
resulting in the expected limits on the fiducial cross-section times branching ratio to dileptons being
(4--7)\% weaker than those without systematic uncertainties.
As all studied signal spin hypotheses (0, 1, 2) have efficiency values which are consistent
within 4$\%$, the limits shown above can be used for reinterpretation of models with such new resonances.

\begin{table}[htp]
\begin{center}
\caption{Observed and expected $95\%$ CL lower limits on $m_{Z'}$ for three $Z'$ gauge boson models, quoted to the nearest $100\,\GeV$ in the $ee$ and $\mu\mu$ channels as well as their combination ($\ell\ell$).}
\label{table:massLimits}
\begin{tabular}{l|cc|cc|cc}
\hline
& \multicolumn{6}{c}{Lower limits on $m_{Z'}$ [\TeV]} \\
Model & \multicolumn{2}{c|}{$ee$} & \multicolumn{2}{c|}{$\mu\mu$} & \multicolumn{2}{c}{$\ell\ell$} \\
& obs & exp & obs & exp & obs & exp \\
\hline
$Z'_\psi$         & $4.1$ & $4.3$ & $4.0$ & $4.0$ & $4.5$ & $4.5$ \\
$Z'_\chi$         & $4.6$ & $4.6$ & $4.2$ & $4.2$ & $4.8$ & $4.8$ \\
$Z'_\mathrm{SSM}$ & $4.9$ & $4.9$ & $4.5$ & $4.5$ & $5.1$ & $5.1$ \\
\hline
\end{tabular}
\end{center}
\end{table}

The generic cross-section limits at $\Gamma/m = 0.5\%$, $1.2\%$ and $3.0\%$ are compared with the model
predictions of $Z'_\psi$, $Z'_\chi$ and $Z'_\mathrm{SSM}$, respectively, to
obtain mass limits. The cross-section values for the model predictions are obtained in the fiducial volume, for compatibility with the definition of the generic signal model.
Mass limits are calculated as the intersection between the expected and observed limits with the model prediction.
Table \ref{table:massLimits} lists the mass limits for the three tested models in all three channels.
These exceed previously reported results~\cite{EXOT-2016-05} by 500--$800\,\GeV$.
 
The generic cross-section limits shown in Figure~\ref{fig:limit}
are smoothly interpolated via Delaunay triangulation~\cite{Delaunay}
to produce limits in between the tested widths.
The results are converted into exclusion contours
in the HVT model coupling space presented in Figure~\ref{fig:hvt},
where $g_\ell$, $g_q$ and $g_h$ correspond to the coupling strengths
between the triplet field and the lepton, quark and Higgs and vector-boson fields,
respectively. In the tested $\{g_q,g_\ell\}$ plane the relative width always remains below $10\%$,
and in the $\{g_h,g_f\}$ plane ($g_f\equiv g_\ell =g_q$) it only exceeds $10\%$
in regions ($|g_f|> 0.9$ and $|g_h|> 2.5$)
well outside the limit contours. The observed limits can be compared with
the limits obtained for the combination of the $\ell\ell$ and $\ell\nu$ channels in Ref.~\cite{EXOT-2017-31} (provided in brackets):
for $g_h=0$ and $m_{Z'_\mathrm{HVT}}=3\,\TeV$, $4\,\TeV$ and $5\,\TeV$
the $|g_f|$ values above 0.07~(0.06), 0.23~(0.15) and 0.49~(0.42) are excluded at 95$\%$ CL, respectively.
The resulting dilepton-only limits are slightly
weaker than those for the $\ell\ell$ and $\ell\nu$ channels combined,
even with a four times larger dataset, because of the higher $W'_\mathrm{HVT}\rightarrow\ell\nu$
cross-section in this model.
 
\begin{figure}[!htpb]
\centering
\subfloat[][]{{\includegraphics[width=0.45\textwidth]{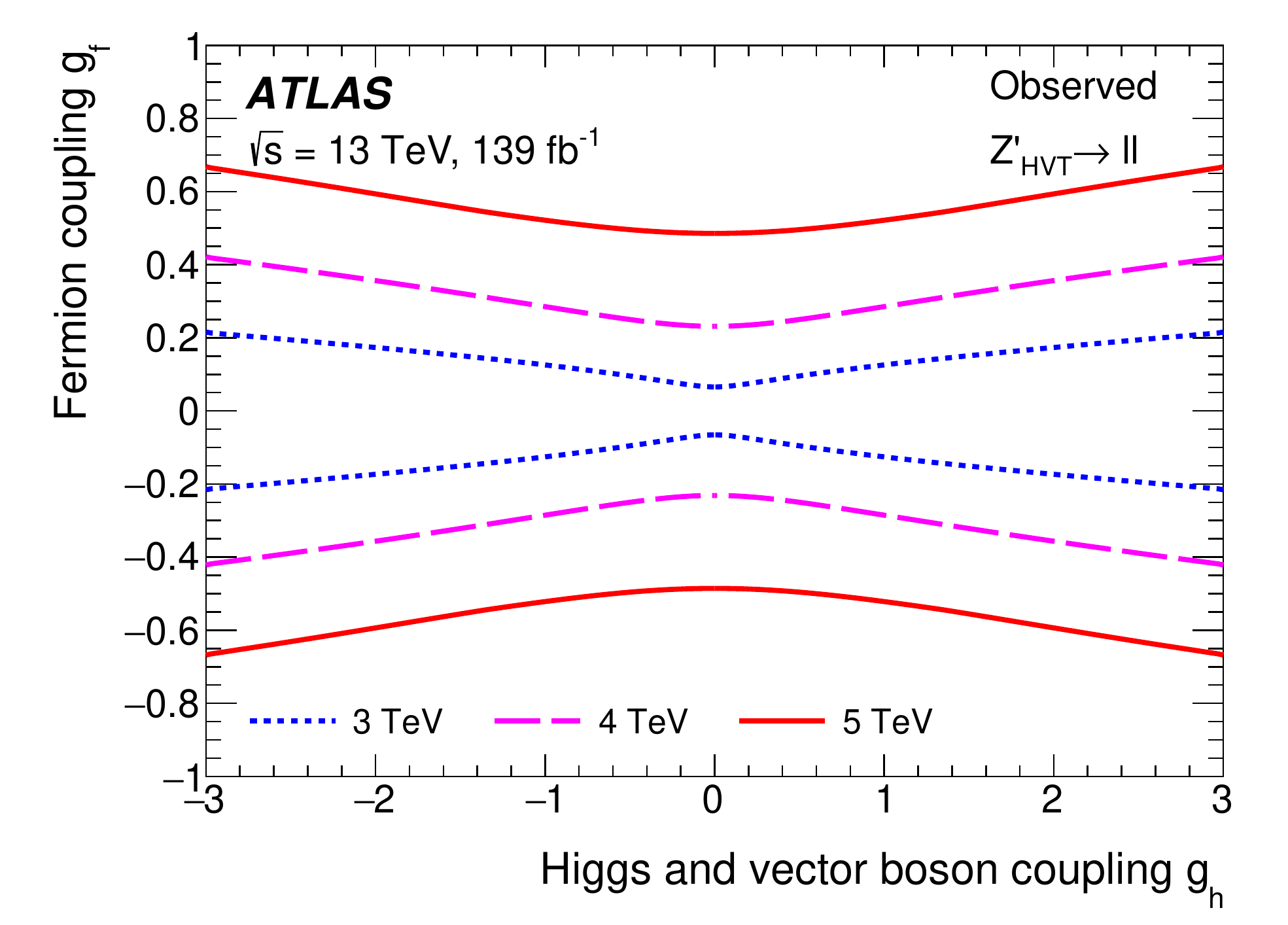}}}
\subfloat[][]{{\includegraphics[width=0.45\textwidth]{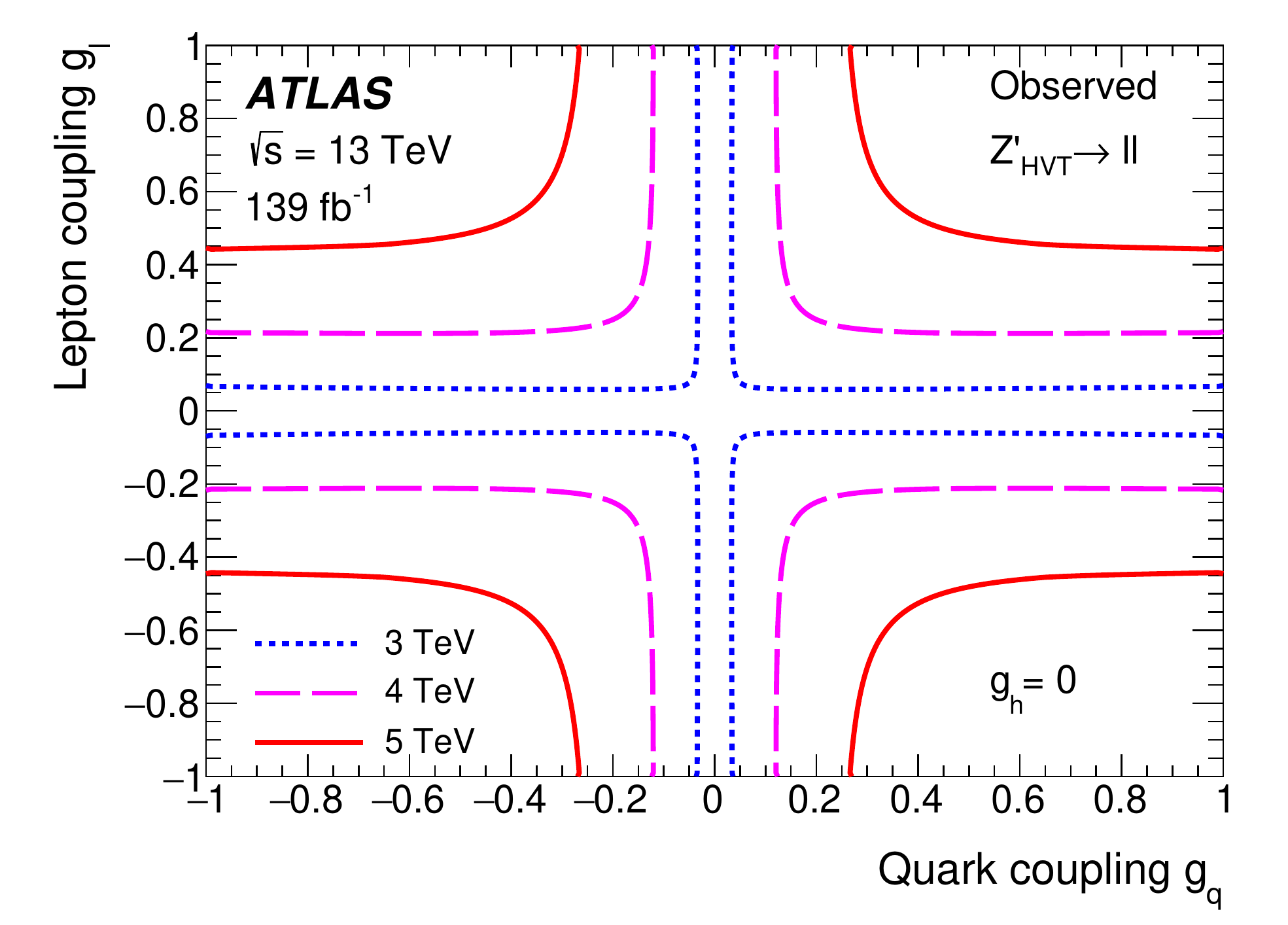}}}
\caption{
Observed 95$\%$ exclusion contours in the HVT parameter space (a) $\{g_h,g_f\}$ with $g_f\equiv g_\ell=g_q$
and (b) $\{g_q,g_\ell\}$ with $g_h$ set to zero,
for resonance masses of 3, 4, and $5\,\TeV$ for the dilepton channel. The area outside the curves is excluded.
}
\label{fig:hvt}
\end{figure}
A complete set of tables and figures (including additional results for the dielectron
and dimuon channels) are available at the Durham HepData repository~\cite{HEPdata}.
 
\section{Conclusions}
\label{sec:conclusion}
The ATLAS detector at LHC is used to search for new
resonances with mass larger than $250\,\GeV$ decaying into a pair of electrons or muons
in  $139\,$\ifb{} of proton--proton collision data at a centre-of-mass energy $\sqrt{s}=13\,\TeV$.
A functional form is fitted to the dilepton invariant-mass distribution in data events
to model the contribution from background processes.
A generic signal shape is used to determine the significance of observed deviations from the
background estimate. No significant deviation is observed.
Limits are set on the fiducial cross-section times branching ratio to dielectrons and dimuons
for generic resonances with a relative natural width in the range of zero
to 10$\%$. These limits are shown to be applicable to spin-0, spin-1 and spin-2 signal
hypotheses. Limits on the Heavy Vector Triplet model couplings and on the masses of vector resonances are inferred.
In particular, the results imply a lower limit of $4.5\, (5.1)\,\TeV$ on $m_{Z'}$
for the $Z'_{\psi}$ ($Z'_\mathrm{SSM}$) boson at 95$\%$ confidence level.
These are the most stringent limits to date.
 
\section*{Acknowledgements}
 
 
We thank CERN for the very successful operation of the LHC, as well as the
support staff from our institutions without whom ATLAS could not be
operated efficiently.
 
We acknowledge the support of ANPCyT, Argentina; YerPhI, Armenia; ARC, Australia; BMWFW and FWF, Austria; ANAS, Azerbaijan; SSTC, Belarus; CNPq and FAPESP, Brazil; NSERC, NRC and CFI, Canada; CERN; CONICYT, Chile; CAS, MOST and NSFC, China; COLCIENCIAS, Colombia; MSMT CR, MPO CR and VSC CR, Czech Republic; DNRF and DNSRC, Denmark; IN2P3-CNRS, CEA-DRF/IRFU, France; SRNSFG, Georgia; BMBF, HGF, and MPG, Germany; GSRT, Greece; RGC, Hong Kong SAR, China; ISF and Benoziyo Center, Israel; INFN, Italy; MEXT and JSPS, Japan; CNRST, Morocco; NWO, Netherlands; RCN, Norway; MNiSW and NCN, Poland; FCT, Portugal; MNE/IFA, Romania; MES of Russia and NRC KI, Russian Federation; JINR; MESTD, Serbia; MSSR, Slovakia; ARRS and MIZ\v{S}, Slovenia; DST/NRF, South Africa; MINECO, Spain; SRC and Wallenberg Foundation, Sweden; SERI, SNSF and Cantons of Bern and Geneva, Switzerland; MOST, Taiwan; TAEK, Turkey; STFC, United Kingdom; DOE and NSF, United States of America. In addition, individual groups and members have received support from BCKDF, CANARIE, CRC and Compute Canada, Canada; COST, ERC, ERDF, Horizon 2020, and Marie Sk{\l}odowska-Curie Actions, European Union; Investissements d' Avenir Labex and Idex, ANR, France; DFG and AvH Foundation, Germany; Herakleitos, Thales and Aristeia programmes co-financed by EU-ESF and the Greek NSRF, Greece; BSF-NSF and GIF, Israel; CERCA Programme Generalitat de Catalunya, Spain; The Royal Society and Leverhulme Trust, United Kingdom.
 
The crucial computing support from all WLCG partners is acknowledged gratefully, in particular from CERN, the ATLAS Tier-1 facilities at TRIUMF (Canada), NDGF (Denmark, Norway, Sweden), CC-IN2P3 (France), KIT/GridKA (Germany), INFN-CNAF (Italy), NL-T1 (Netherlands), PIC (Spain), ASGC (Taiwan), RAL (UK) and BNL (USA), the Tier-2 facilities worldwide and large non-WLCG resource providers. Major contributors of computing resources are listed in Ref.~\cite{ATL-GEN-PUB-2016-002}.
 

\printbibliography

\clearpage 
 
\begin{flushleft}
{\Large The ATLAS Collaboration}

\bigskip

G.~Aad$^\textrm{\scriptsize 101}$,    
B.~Abbott$^\textrm{\scriptsize 128}$,    
D.C.~Abbott$^\textrm{\scriptsize 102}$,    
O.~Abdinov$^\textrm{\scriptsize 13,*}$,    
A.~Abed~Abud$^\textrm{\scriptsize 70a,70b}$,    
K.~Abeling$^\textrm{\scriptsize 53}$,    
D.K.~Abhayasinghe$^\textrm{\scriptsize 93}$,    
S.H.~Abidi$^\textrm{\scriptsize 167}$,    
O.S.~AbouZeid$^\textrm{\scriptsize 40}$,    
N.L.~Abraham$^\textrm{\scriptsize 156}$
% ...
\bigskip
\\

% ...
$^{13}$Institute of Physics, Azerbaijan Academy of Sciences, Baku; Azerbaijan.\\
% ...
$^{40}$Niels Bohr Institute, University of Copenhagen, Copenhagen; Denmark.\\
% ...
$^{53}$II. Physikalisches Institut, Georg-August-Universit\"{a}t G\"ottingen, G\"ottingen; Germany.\\
% ...
$^{70}$$^{(a)}$INFN Sezione di Pavia;$^{(b)}$Dipartimento di Fisica, Universit\`a di Pavia, Pavia; Italy.\\
% ...
$^{93}$Department of Physics, Royal Holloway University of London, Egham; United Kingdom.\\
% ...
$^{101}$CPPM, Aix-Marseille Universit\'e, CNRS/IN2P3, Marseille; France.\\
$^{102}$Department of Physics, University of Massachusetts, Amherst MA; United States of America.\\
% ...
$^{128}$Homer L. Dodge Department of Physics and Astronomy, University of Oklahoma, Norman OK; United States of America.\\
% ...
$^{156}$Department of Physics and Astronomy, University of Sussex, Brighton; United Kingdom.\\
% ...
$^{167}$Department of Physics, University of Toronto, Toronto ON; Canada.\\
% ...
$^{a}$ Also at Centre for High Performance Computing, CSIR Campus, Rosebank, Cape Town; South Africa.\\
$^{b}$ Also at CERN, Geneva; Switzerland.\\
% ...
$^{*}$ Deceased

\end{flushleft}


\end{document}